\documentclass[conference]{IEEEtran}
\usepackage{filecontents}
\usepackage[noadjust]{cite}

\usepackage{tikz}
\usetikzlibrary{shapes,arrows}

\title{Efficient Energy Harvesting in Wireless Sensor Networks of Smart Grid}

\author{\IEEEauthorblockN{Md. Enamul Haque, Ahmad Shawahna, and Dr. Uthman Baroudi}
\IEEEauthorblockA{Department of Computer Engineering\\
King Fahd University of Petroleum and Minerals, Dhahran-31261, KSA\\
\{g201204920, g201206920, ubaroudi\}@kfupm.edu.sa}}

\begin{document}
\maketitle

\begin{abstract}
Smart grids are becoming ubiquitous in recent time. With the progress of automation in this arena, it needs to be diagnosed for better performance and less failures. There are several options for doing that but we have seen from the past research that using Wireless Sensor Network (WSN) as the diagnosis framework would be the most promising option due to its diverse benefits. Several challenges such as effect of noise, lower speed, selective node replacement, complexity of logistics, and limited battery lifetime arise while using WSN as the framework. Limited battery lifetime has become one of the most significant issues to focus on to get rid of it. This article provides a model for replenishing the battery charge of the sensor nodes of wireless sensor network. We will use the model for sensor battery recharging in an efficient way so that no nodes become out of service after a while. We will be using mobile charger for this purpose. So, there may be some scope for improving the recharge interval for the mobile charger as well. This will be satisfied using optimum path calculation for each time the charger travels to the nodes. Our main objectives are to maximize the nodes battery utilization, distribute power effectively from the energy harvester, and minimize the distance between power source and cluster head. The simulation results show that the proposed approach successfully maximizes the utilization of the nodes battery while minimizes the waiting time for the sensor nodes to get recharged from the energy harvester. 
\end{abstract}

\IEEEpeerreviewmaketitle

\section{Introduction}
Wireless sensor networks (WSN) are promising option for monitoring smart grids if that is provided with low cost and large geographic area coverage. There may arise some issues related to power consumption of each sensor nodes for their varying rate of energy consumption. Thus the main power sink need to measure the remaining power level of these nodes and recharge them after certain intervals. There is another solution for the extreme environmental states where the battery life should be longer than usual. In this paper we will focus on the efficient distribution of power for each of the nodes via an energy harvester that will learn about the smart grid environment in course of time with continuous monitoring. We will introduce the detail steps in the following sections.

WSNs are becoming ubiquitous in smart grid environments in recent years. There are some research issues of WSNs in the capacity, routing strategy, and delay model, application etc. One of the significant issues is with the battery life time of the sensor nodes. If we use some static approach to recharge the nodes, there will be a lot of wastage of energy considering large smart grid environment. Thus the energy provider or the sink should have some strategy based on the input from the grids sensor nodes, so that it can emit the power in an efficient way. We will be using one mobile charger for this purpose.

The energy harvester, \textit{$E_h$} will consider these three things while exploring the grid environment to disseminate the energy to the sensors:\\
1. Remaining energy level of sensor nodes.\\
2. Remaining energy level of the harvesters.\\
3. Shortest path from the harvester to the nodes.

We need to consider several situations and stipulate algorithms accordingly to meet the above criteria and provide an energy saving mode in the grid environment. In the following sections we will discuss briefly about the grid monitoring through WSN nodes, issues and probable solutions with algorithms. These will enhance the overall power consumption performance of the grid.

In a wireless sensor network environment, landmark refers to the points from where it is suitable for the harvester to emit power to the nodes. Several energy harvesters will be deployed in the grid after monitoring the power level and the quickest path from the harvester to the cluster of nodes. The harvester will move from one landmark to another within a cluster based on the requirement of the nodes and on its own power level. If it cannot suffice the nodes with energy of a certain cluster, it will send the information to its neighboring harvesters from the adjacent clusters.

Energy harvester visits each cluster based on the power consumption, so the energy replenishment issue is taken cared after each round. This decision is made on the fly from the input from base station. We assume the base station stays at the center of the network and the power source for the harvester is outside the network coverage. So, we need to consider the round trip delay between each round to maximize the energy utilization among the nodes.

The rest of this paper is organized as follows. In section II, we mention some related work. In section III, we cover the RF energy harvesting ideas briefly. In section IV, we present the energy consumption model and the technical setup for the simulation of the model. Section V concludes this paper.

\section{Related Work}
There are several works related to energy harvesting for sensor nodes from previous authors. They proposed different approaches for the solution. We will be having some of their ideas to get the difference from our approach.

In this paper, \cite{erol2012suresense} the author explained about achieving  timely and efficient charging of the sensor nodes. They proposed one technique which includes Mobile wIreless Charger RObots (MICRO) for replenishing the battery of the sensor nodes. They introduced the concept regarding landmark which refers to a location in the network from where all the nodes of a cluster can receive power with minimum cost. They used minimum number of landmarks with location and energy replenishment requirements of the sensors. Their research showed two options: landmark selection, and reach that landmark using shortest path possible. This model will for sure help to reduce some energy consumption. Specific energy consumption model for the sensors were not mentioned in the paper though.

\textbf{Sinha} mentioned about \cite{sinha2001dynamic} improving the energy efficiency of the sensor nodes by OS directed power management technique. They considered the state transitions of the sensor nodes. As we know all the nodes in a cluster are not always busy sending and receiving data. So, there are nods which are idle in each moment. So, what they proposed is similar with operating system behavior. OS goes into sleep mode after some idle time. Similarly sensor nodes will become inactive after some threshold inactivity. They also mentioned that this sleep and awake state transformation would take additional energy which maybe more than if the node would have been alive. So, there maybe some improvement areas in this regard.

\textbf{Kantarci} \cite{erol2012drift} analyzed about maximizing the energy consumption for the high priority nodes. They determined how to select high and low priority nodes using DRIFT (Differentiated RF Power Transmission) scheme. The features were power reception energy per (high/low) priority nodes, and path traversal efficiency. Then they compared their result with their previous paper tilted SURESENSE and found that previous one exploited better result than DRIFT. The priority scheme needs to be updated along with the energy consumption model as we noticed.

\textbf{Shi} et al. \cite{shi2011renewable} studied an optimization problem on maximizing the battery lifetime of the nodes which is similar to our purpose. But, we approach with different energy model and found that it provides better result. They introduced the wireless charging vehicle (WCV) which is similar in nature like the MICROs used in \cite{erol2012suresense}. They optimized the vacation time and the travelling path for the charger using shortest Hamiltonian cycle. They provided one near optimal solution to prove its performance.
 
\textbf{Gungor} et al. \cite{gungor2010opportunities} emphasized on the generation, delivery and utilization of the electrical power systems via smart grid. Their aim was to explore the opportunities and challenges faced in a wireless sensor network enabled smart grid where. They showed comprehensive experimental study on the statistical characterization of the wireless channel in different electrical power system environments. They also performed some field tests on IEEE 802.15.4-complaint wireless sensor nodes in real world to measure noise, channel characteristics, and attenuation in frequency band. They presented some insights about design decision and trade-offs for WSN empowered smart grid applications.

\textbf{Kantarci} et al. \cite{erol2011wireless} proposed one energy management scheme for residential usage called iHEM (in home energy management). They focus on the demand supply balance and electricity expense reduction. They also compared between iHEM and OREM (Optimization based residential energy management) schemes and found it performs almost near to OREM. They considered local energy generation capability, priority and real time pricing while performance evaluation. Finally, they show that the packet delivery ration, delay and jitter of wireless sensor home area network (WSHAN) improve when packet size increases.

\textbf{Chen} et al. \cite{he2011energy} studied WRSN (Wireless Rechargeable Sensor Network) built from WISP (Wireless Identification and Sensing Platform) and RFID (Radio Frequency Identification Device). They mentioned that the WISP tags can harvest energy from receiving the energy from the readers. They attended the issue of continuous operation for these WISP tags in indoor environment and referred to as \textit{energy provisioning} issue. Their analysis reveals that their deployment methods reduce the number of readers significantly compared to those assuming traditional coverage models.

\textbf{Sudevalayam} et al. \cite{sudevalayam2011energy} studied the implications of recharge opportunities on sensor node operation and design of sensor network solutions. They focuses on the battery power nodes and how these nodes meets the goals of lifetime, cost, data sensing reliability, transmission coverage, energy harvesting, ambient energy conversion to electrical energy etc. They surveyed different aspects of energy harvesting sensor system architecture and their storage technologies with examples.

In this paper we extend the work of Shi et al. \cite {shi2011renewable} to maximize the life time of wireless sensor nodes with novel energy consumption model.

\section{Wireless Power Transmission and Sensor Nodes Lifetime}

Wireless sensor networks are cost efficient as they need very less energy to transmit data over the wireless network. Sensors participate to use low power media access control protocol for information sharing. The nodes replenish their batteries with replacement or recharging. It becomes cost effective if the batteries are replaced very often, so recharging becomes realistic option. 

Wireless recharging technology becomes more sophisticated in case of radio frequency transmission. The quality of energy transfer hinders from different issues. For example, interference and noise can make the energy less transmission from the actual. Thus the energy modeling should take into these challenges and keep some resource allocated. On the other hand sensor nodes lifetime depends on the battery or the self energy source. We could model the recharging option using solar energy, but it is very likely that there will be time when the solar energy is not available. Then the smart grid can not be monitored due to inactive sensors. Thus the best option would be the mix of solar and RF energy transfer.

Though the hybrid energy transfer is a feasible option, the implementation is complex due to the hardware requirement and cost. Thus we are considering only RF energy transmission throughout our article. The energy used by a sensor node consists of energy consumed by receiving, transmitting, listening for message on the radio channel, sampling data and sleeping \cite {polastre2004versatile} .

In wireless sensor network environment, it is also very likely that some nodes are prone to sit idle while others are highly active for information transmission \cite {jeong2007empirical}. So, some nodes can act as power source within the cluster considering the on demand mechanism. So, this model is also quite complex to implement due to its dynamic behavior. 

Recently, WSNs are becoming the choice for supporting data communication in industrial applications. For example, HVAC and control mechanism which were previously based on wired link. To implement in this environment we need to provide the sensors the capability to transfer information in multi hop communication method.\cite {schmidt2007energy} suggested one energy modelling using finite state machine implementation in the node.

We have described our simulation which implicitly uses TDMA to make the slot allocation for the nodes while transferring and receiving data from the base station. We considered several set of nodes (50,100,150) and made the simulation based on that. We observed that the network never becomes inactive as in every round the sensors are gaining charge.

\section{Experiment}

We have used the below hardware and software configuration to run our simulation. All the nodes communicate to the base station to get the status update regarding the energy level. The energy harvester which acts mobile also gets the information from the base station. Thus the synchronization is being done via the BS (Base Station) between harvester and the nodes.\\

\begin{enumerate}
\item Operation System: Fedora 
\item Simulator: NS 2
\item Additional plugin for NS2: LEACH
\item Compiler: TCL
\item Number of nodes in the network: 50 /100 /150 nodes.
\item Number of clusters in the network: 5\% of the total nodes.
\item Initial energy for each node: 2 Joule.
\item Time for each round: 20 second.
\end{enumerate}

We assumed the nodes in the network are distributed uniformly and the energy for each node is calculated at the beginning of each round. Each node gets energy using the below model. Distance is being considered, so the remote nodes from the power source will get less energy and the closer ones will get more energy at each round. This is one of the bottleneck, which is resolved by using both distance and the remaining energy level of the node.

\begin{figure}[t]
\centering
\includegraphics [scale=0.5]{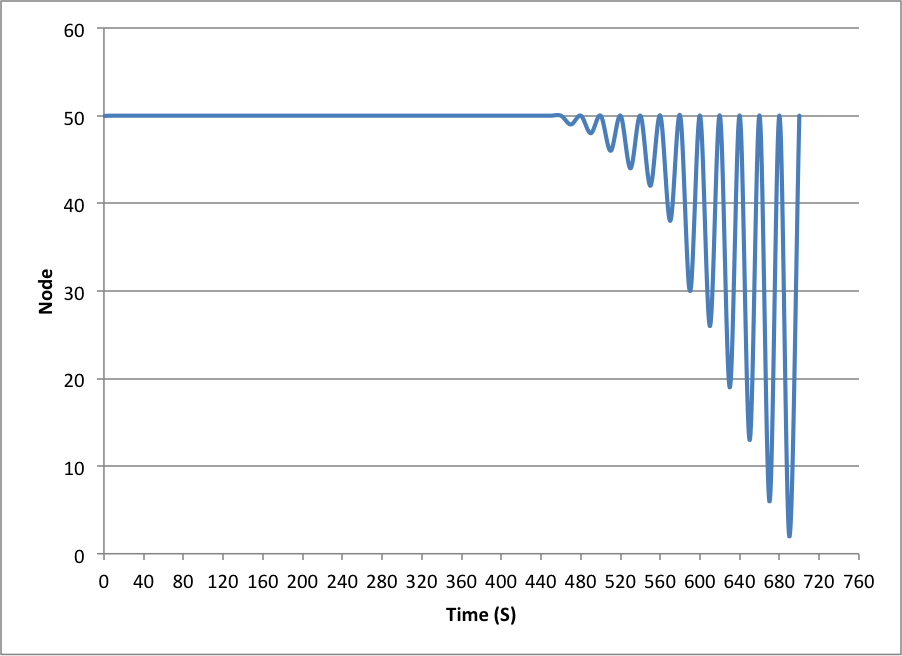}
\caption{Time Vs. No of Alive Nodes (Node in the network=50)}
\label{fig_sim}
\end{figure}

\begin{figure}[t]
\centering
\includegraphics [scale=0.5]{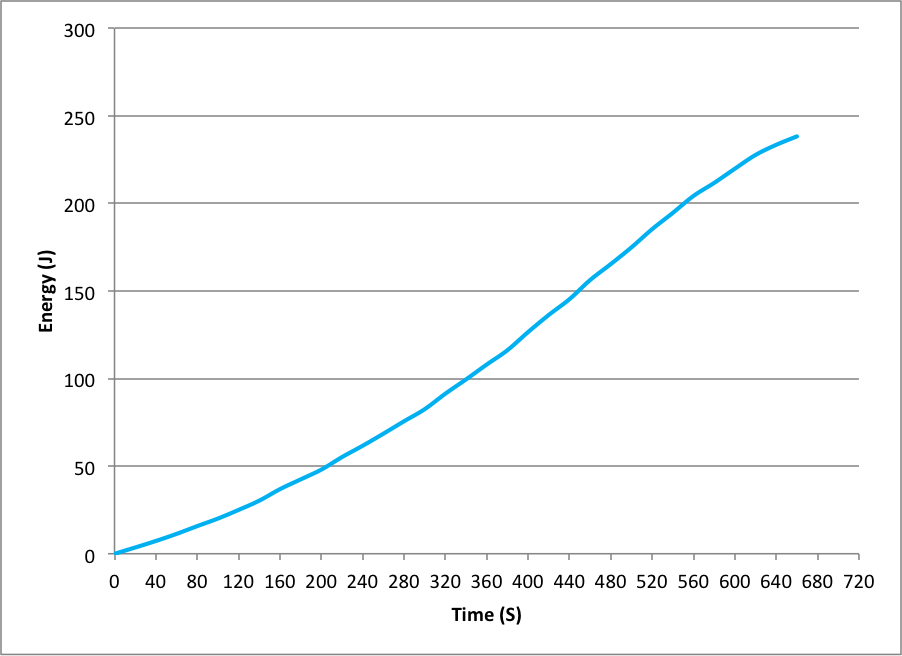}
\caption{Time Vs. Total Consumed Energy (50 Nodes)}
\label{fig_sim}
\end{figure}

The below equation provides the simple representation of what we implemented in LEACH using NS2 simulator.

\begin {equation}
E_n = {E_c} + \frac{E_h}{d^2}
\end {equation}

Where $E_n$ = New energy for the node.\\
$E_c$ = Current Energy in the node.\\
$E_h$ = Harvester Energy. (Depends on the consumed energy in the Energy for each H cluster).\\
$d$ = Distance between the node and the harvester.\\

The area we considered for the simulation is 100 by 100 square meter. The nodes are spread over the area uniformly. This is taken cared of by LEACH. Nodes are distributed in each cluster,meaning some nodes become group and one leader from that group is selected as cluster head. The clustering process is done by the LEACH as well depending on the required energy of each nodes. 

The energy harvester communicates with the BS to know the required amount of energy by the network. The harvester receives the energy from the power station based on the amount of energy consumed in the network in the previous round. Then harvester gets the information about which cluster to visit first and how much energy to harvest for that cluster from the base station.

Harvester reaches to the cluster head location of each cluster and stay for a specified time to harvest energy. Sensor nodes which are near the cluster head gain more energy than the far ones. This process is repeated until all clusters are visited by the harvester.  

\begin{figure}
\centering
\includegraphics [scale=0.5]{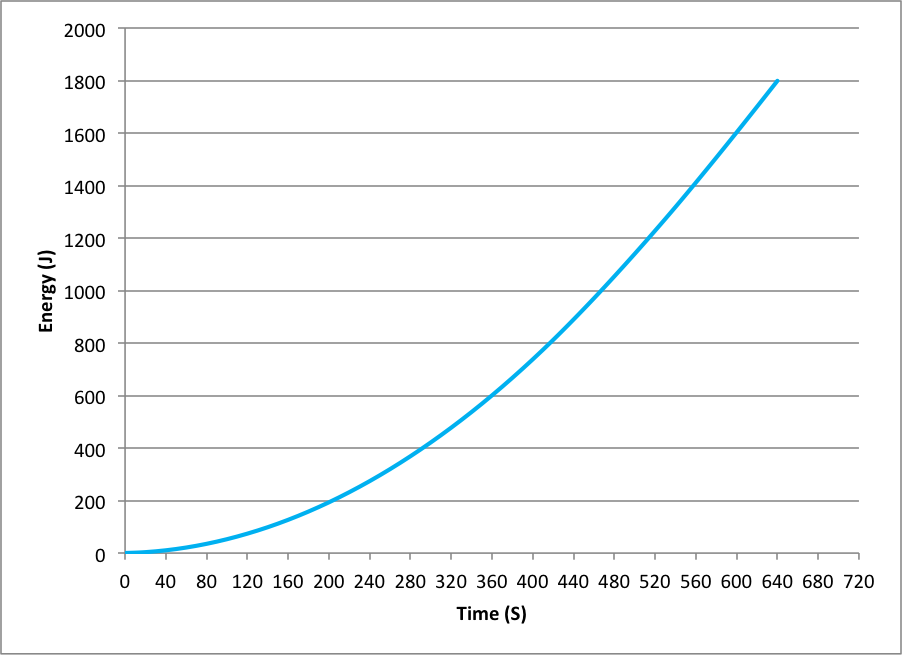}
\caption{Time Vs. Total Harvested Energy (50 Nodes)}
\label{fig_sim}
\end{figure}

\begin{figure}
\centering
\includegraphics [scale=0.5]{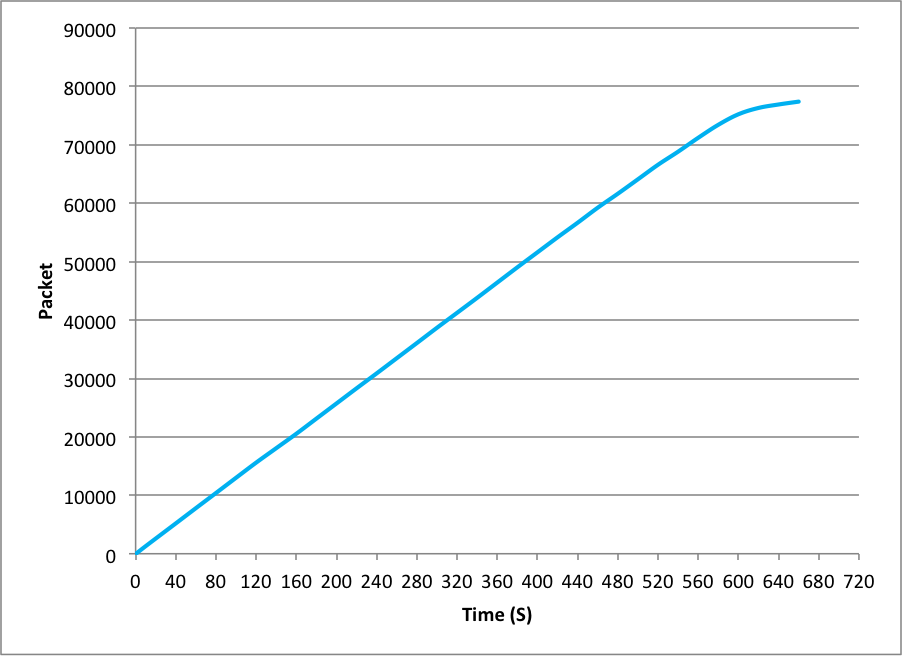}
\caption{Time Vs. Total Data Received (50 Nodes)}
\label{fig_sim}
\end{figure}

\begin{figure}[b]
\centering
\includegraphics [scale=0.5]{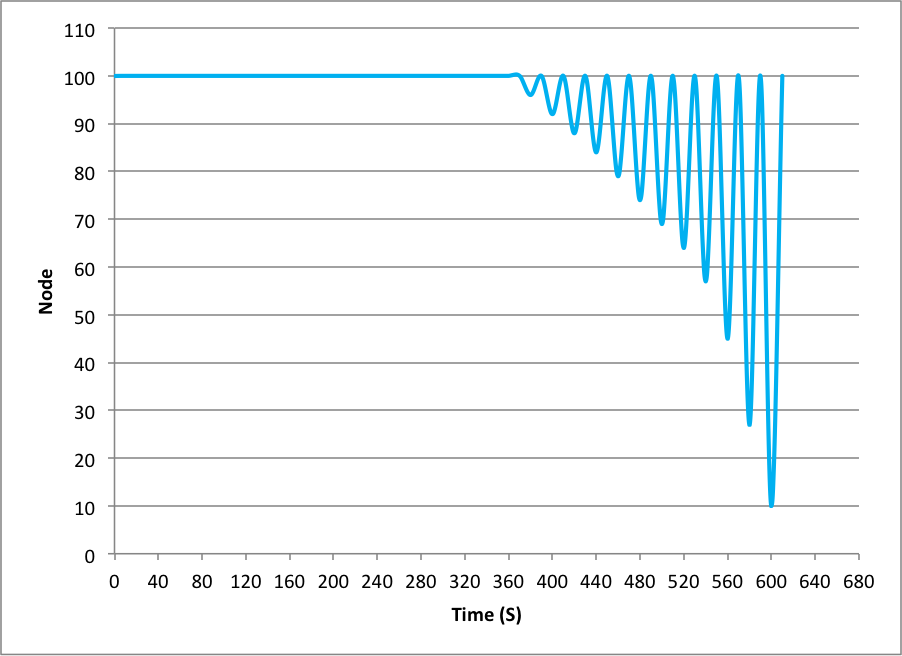}
\caption{Time Vs. No of Alive Nodes (Node in the network=100)}
\label{fig_sim}
\end{figure}

\begin{figure}[t]
\centering
\includegraphics [scale=0.5]{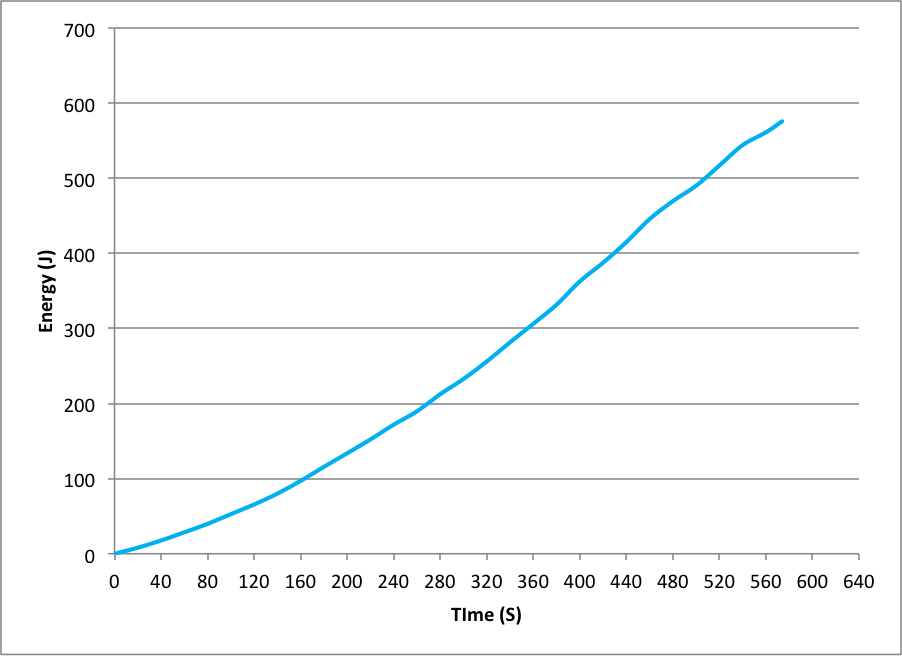}
\caption{Time Vs. Total Consumed Energy (100 Nodes)}
\label{fig_sim}
\end{figure}

\begin{figure}[t]
\centering
\includegraphics [scale=0.5]{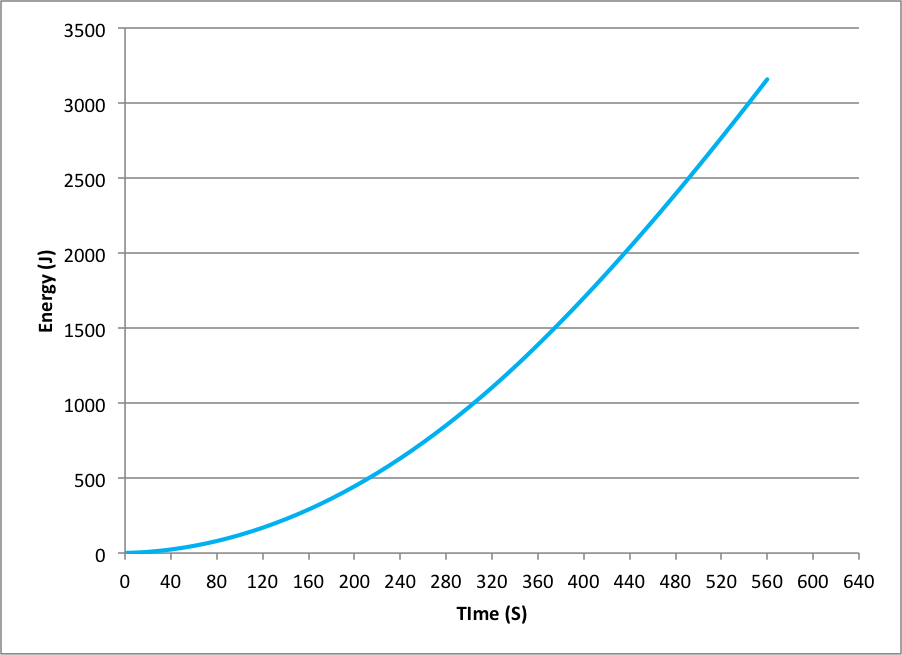}
\caption{Time Vs. Total Harvested Energy (100 Nodes)}
\label{fig_sim}
\end{figure}

\begin{figure}[b]
\centering
\includegraphics [scale=0.5]{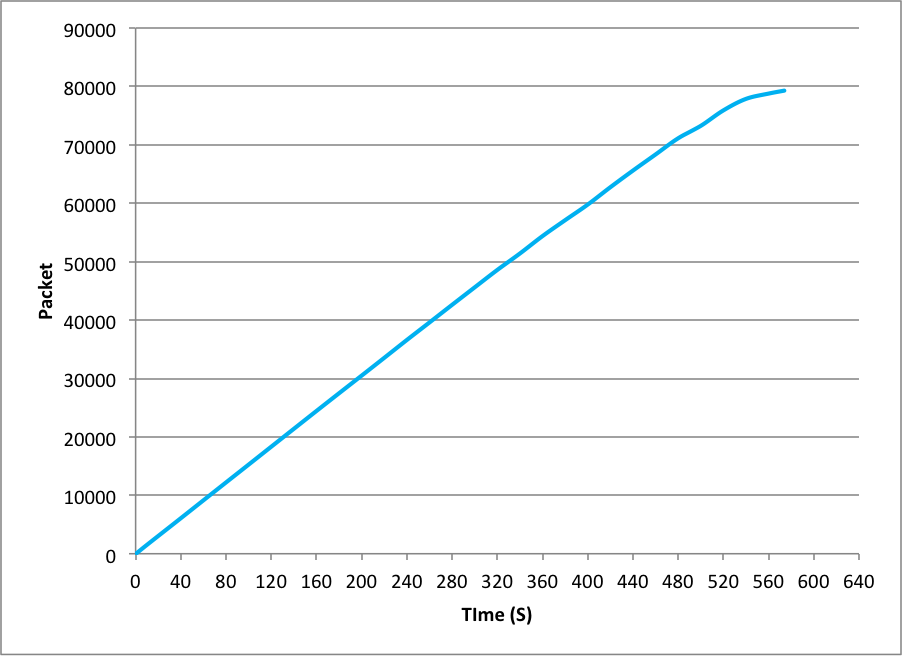}
\caption{Time Vs. Total Data Received (100 Nodes)}
\label{fig_sim}
\end{figure}

\begin{figure}[t]
\centering
\includegraphics [scale=0.5]{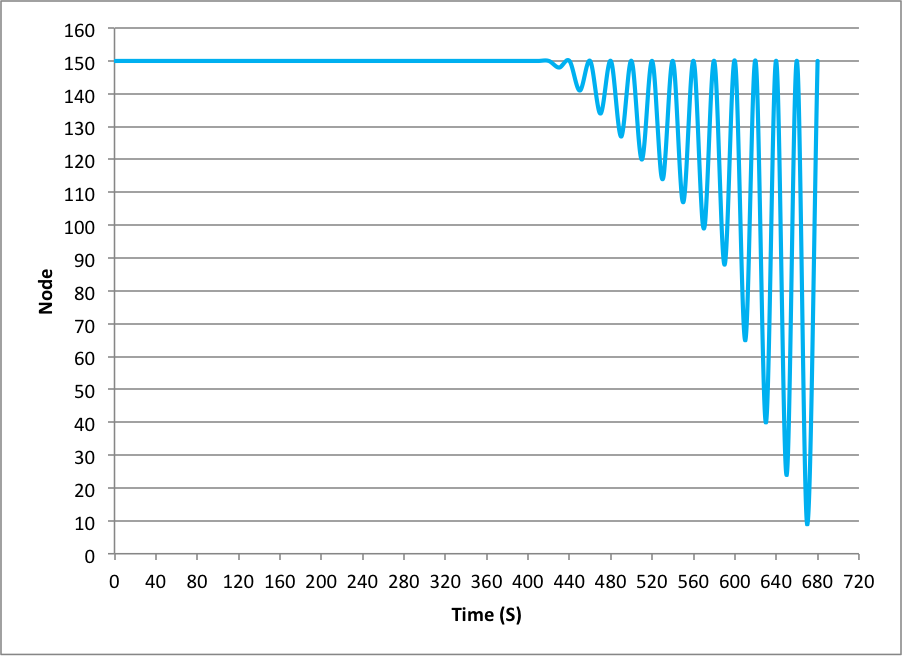}
\caption{Time Vs. No of Alive Nodes (Node in the network=150)}
\label{fig_sim}
\end{figure}

\section{Conclusion}

Smart grid requires robust and constant monitoring to ensure quality of service (QoS). Wireless sensor networks becomes handy considering different aspects in this purpose. This article provides a model for replenishing the battery charge of the sensor nodes of wireless sensor network. The proposed model employs mobile charger for sensor battery recharging. The mobile charger uses optimum path calculation for each time the charger travels to the nodes. The simulation results show that the proposed approach successfully maximizes the utilization of the nodes battery while minimizes the waiting time for the sensor nodes to get recharged from the energy harvester. 

\section{Future Work}

We noticed that after some time the number of alive nodes decreases drastically and then reaches to its original state. So, this increases the frequency of the travelling of the harvester, thus increases in power consumption. So, this issue should be taken cared. We will try to address this in our future work.

\bibliography{main}
\bibliographystyle{IEEEtran}
\end{document}